\documentclass[conference]{IEEEtran}
\IEEEoverridecommandlockouts
\pdfoutput=1
\usepackage{cite}
\usepackage{amsmath,amssymb,amsfonts}
\usepackage{algorithmic}
\usepackage{graphicx}
\usepackage{textcomp}
\usepackage{xcolor}
\usepackage{multirow}
\usepackage{enumitem}
\usepackage{algorithm}
\usepackage{comment}
\usepackage{tikz}
\usepackage{hyperref}
\usepackage{lipsum}

\newcommand\copyrighttext{%
  \footnotesize \textcopyright 2019 IEEE. Personal use of this material is permitted. Permission from IEEE must be obtained for all other uses, in any current or future media, including reprinting/republishing this material for advertising or promotional purposes, creating new collective works, for resale or redistribution to servers or lists, or reuse of any copyrighted component of this work in other works.}
\newcommand\copyrightnotice{%
\begin{tikzpicture}[remember picture,overlay]
\node[anchor=south,yshift=10pt] at (current page.south) {\fbox{\parbox{\dimexpr\textwidth-\fboxsep-\fboxrule\relax}{\copyrighttext}}};
\end{tikzpicture}%
}

\newcommand{\sign}{\operatorname{sign}}
\def\BibTeX{{\rm B\kern-.05em{\sc i\kern-.025em b}\kern-.08em
    T\kern-.1667em\lower.7ex\hbox{E}\kern-.125emX}}
\begin{document}

\title{TDMR Detection System with Local Area
Influence Probabilistic {\em a Priori} Detector
}

\author{\IEEEauthorblockN{1\textsuperscript{st} Jinlu Shen}
\IEEEauthorblockA{\textit{Washington State University} \\
\textit{School of EECS}\\
Pullman, WA 99164-2752 USA \\
jinlu.shen@wsu.edu}
\and
\IEEEauthorblockN{2\textsuperscript{nd} Xueliang Sun}
\IEEEauthorblockA{\textit{Quantamental Technologies LLC} \\
White Plains, NY 10606 USA\\
xueliang.sun@wsu.edu}
\and
\IEEEauthorblockN{3\textsuperscript{rd} Krishnamoorthy Sivakumar}
\IEEEauthorblockA{\textit{Washington State University, School of EECS}\\
Pullman, WA 99164-2752 USA \\
siva@wsu.edu}
\and
\IEEEauthorblockN{4\textsuperscript{th} Benjamin J. Belzer}
\IEEEauthorblockA{\textit{Washington State University, School of EECS}\\
Pullman, WA 99164-2752 USA \\
belzer@wsu.edu}
\and
\IEEEauthorblockN{5\textsuperscript{th} Kheong Sann Chan}
\IEEEauthorblockA{
\textit{Nanjing Institute of Technology} \\
Nanjing, China\\
kheongsann@ieee.org}
\and
\IEEEauthorblockN{6\textsuperscript{th} Ashish James}
\IEEEauthorblockA{\textit{Institute for Infocomm Research (I2R)} \\
\textit{A*STAR}, Singapore \\
ashish\_james@i2r.a-star.edu.sg}
}

\maketitle
\copyrightnotice

\begin{abstract}
We propose a three-track detection system for two dimensional magnetic recording (TDMR) in which a local area influence probabilistic (LAIP) detector works with a trellis-based Bahl-Cocke-Jelinek-Raviv (BCJR) detector to remove intersymbol interference (ISI) and intertrack interference (ITI) among coded data bits as well as media noise due to magnetic grain-bit interactions.
Two minimum mean-squared error (MMSE) linear equalizers with different response targets are employed before the LAIP and BCJR detectors. The LAIP detector considers local grain-bit interactions and passes coded bit log-likelihood ratios (LLRs) to the channel decoder, whose output LLRs serve as {\em a priori} information to the BCJR detector, which is followed by a second channel decoding pass. Simulation results under 1-shot decoding on a grain-flipping-probability (GFP) media model show that the proposed LAIP/BCJR detection system achieves density gains of $6.8\%$ for center-track detection and $1.2\%$ for three-track detection compared to a standard BCJR/1D-PDNP. The proposed system's BCJR detector bit error rates (BERs) 
are lower than those of a recently proposed two-track BCJR/2D-PDNP system by factors of
$(0.55, 0.08)$ for tracks 1 and 2 respectively.

\end{abstract}

\begin{IEEEkeywords}
Two-dimensional magnetic recording, iterative detection and decoding, local area influence probability,
grain-flipping-probability model
\end{IEEEkeywords}

\section{Introduction}
Two dimensional magnetic recording (TDMR) is a promising technology for increasing the areal density of next generation hard disk drives (HDDs) without requiring radical redesign of recording media. 
Proposed generalizations of 1D pattern dependent noise prediction (1D-PDNP, \cite{Kavcic_IT,Moon_JSAC}) to two-track TDMR (such as, e.g., \cite{Yao_ICC15,Wang_Kumar_2016_1,Shi_Barry_2018}) have trellis state cardinality
$4^{(\Delta+I+L)}$, where $\Delta$ is the predictor look-ahead, and $I$ and $L$ are the intersymbol interference (ISI) channel length and predictor order. The complexity grows rapidly with $I+L$, and becomes impractical for more than two tracks. 

Processing three tracks can account for intertrack interference (ITI) from both adjacent tracks to the center track, 
leading to significant density gains, especially at lower track pitches.
Thus, as an alternative to 2D-PDNP, in \cite{Sun2017} we designed a three-track local area influence probabilistic (LAIP) detector for data from a Voronoi magnetic grain model. In initial offline training, the LAIP collects local bit influences on the target bit and discretizes their frequencies into a multidimensional joint probability mass function (PMF), organized as a lookup table (LUT). In detection mode, the LAIP detector searches the LUT for bit influences, and compares the readback signal with estimated overall bit influences to obtain log-likelihood ratios (LLRs). Then LLRs are exchanged between detector and channel decoder until convergence occurs.

In this paper, we incorporate pre-processing of data from a grain-flipping-probability (GFP) media model through minimum mean-squared error (MMSE) linear equalizers, and propose modified LAIP training methods tailored to the pre-processed data. The GFP model is a realistic HDD media model that provides fast and accurate 2D readback waveforms that include effects captured from micromagnetic simulations \cite{GFPchan1}. The GFP model has been validated in previous studies against both spin-stand \cite{Chan2012,Teo2012} and HDD \cite{Chan2015} signals, and an HDD areal density estimate has been made in \cite{Elidrissi2011}.

This paper's novel contributions are: 1) A parallel MMSE filtering and full and partial response (FR and PR) signaling architecture with filters and FR and PR targets optimized for the LAIP and Bahl-Cocke-Jelinek-Raviv (BCJR) detectors respectively; 2) Optimization of the LAIP PMF table structure to account for MMSE filtering and PR signaling; 3) Simulation results on GFP-simulated TDMR waveforms showing up to $6.8\%$ density gains over conventional 1D MMSE/BCJR/PDNP, and bit error rate (BER) reductions by as much as a factor of $0.08$ compared to a recently proposed 2D-PDNP scheme.

This paper is organized as follows. Section~\ref{sec: overview} provides an overview of the proposed LAIP/BCJR TDMR detection system. Section~\ref{sec: laip} describes the LAIP detector, including training and detection algorithms. Section~\ref{sec: tdmr_system} explains the combined LAIP/BCJR TDMR detection system, and section~\ref{sec: simu_results} gives Monte-Carlo simulation results on the GFP model.

\section{TDMR detection system overview}
\label{sec: overview}
\begin{figure}[ht]
\centerline{\includegraphics[width=\columnwidth]{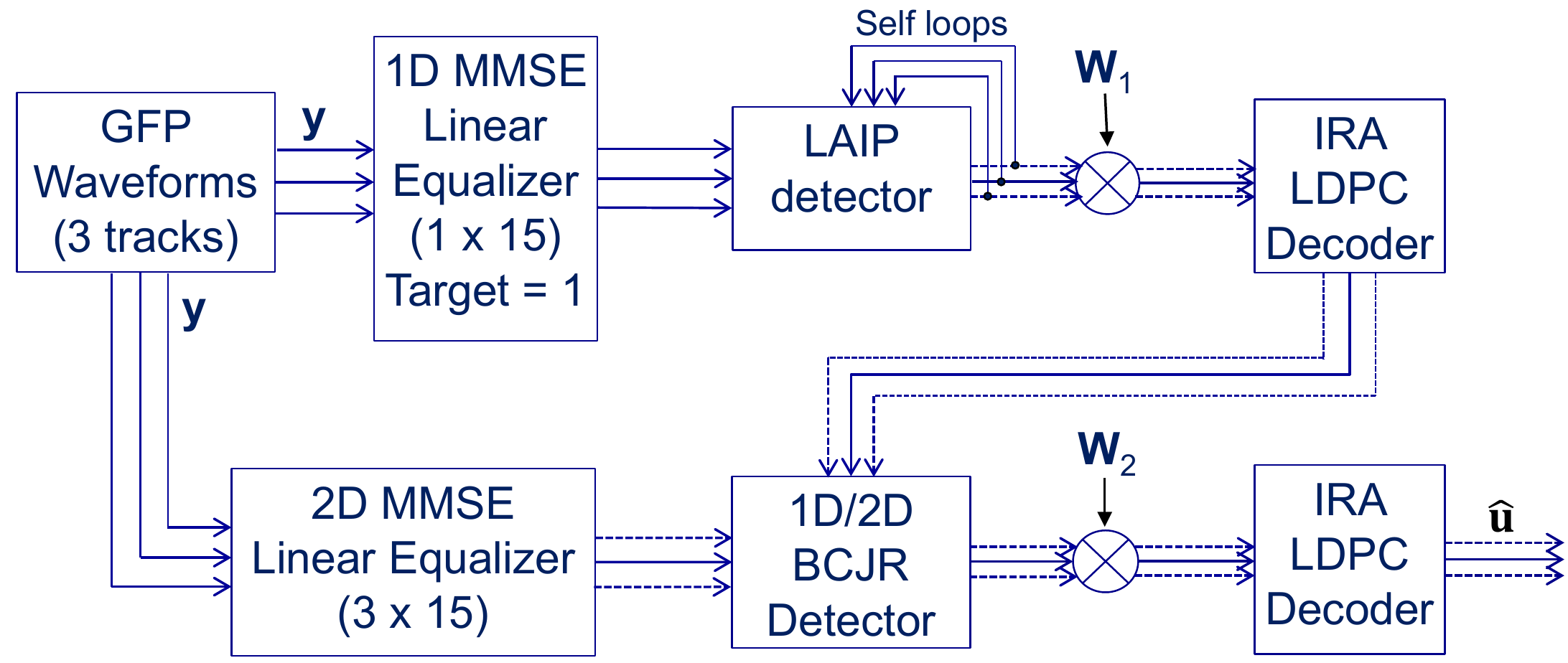}}
\centering
\vspace{-0.1in}
\caption{Block diagram of the LAIP/BCJR TDMR detector. The dotted lines indicate three-track detection.}
\label{fig: Y-eff-turbo-det}
\vspace{-0.1in}
\end{figure}

Fig.~\ref{fig: Y-eff-turbo-det} shows a block diagram of the proposed LAIP/BCJR TDMR detection system. 
The system inputs are three tracks of GFP waveforms. There are two data flow paths. Both paths have an MMSE linear equalizer for pre-processing the data, a detector that reduces the BER by roughly a factor of $1/2$, and a channel decoder. For the MMSE linear equalizers, a linear filter $h$ is applied on the raw GFP readings $y$ in order to minimize the MSE between the filtered output $h*y$ and $g*u$, where $u$ is the block of coded data bits, $g$ is a PR or FR target, and $*$ indicates discrete 1D or 2D convolution. The BCJR detector is a soft-in/soft-out trellis-based detector that uses the PR target and minimizes the symbol error probability. The LAIP detector is a probabilistic model-based detector that uses pre-trained multidimensional joint PMFs to estimate the influence of the surrounding bits on a given target bit in a local area. Unlike PDNP-based detectors, the BCJR detector does not develop its own estimate of the data dependent media noise; instead, the LAIP detector implicitly supplies media noise information through the coded bit LLRs it passes to the first channel decoder and then to the BCJR. Multiplicative weights ${\bf w}_1$ and ${\bf w}_2$ are applied to the LLRs passed by the LAIP and BCJR detectors to the IRA decoders. 

The channel decoders are irregular repeat accumulate (IRA) low density parity check (LDPC) decoders \cite{mceliece}, which employ soft coset decoding based on the known (but randomly distributed) input bits on each track. The two channel decoders are identical, and process the three tracks independently, under the assumption that the information on each track is encoded separately by a single channel encoder using the same code rate for each track.

In Fig.~\ref{fig: Y-eff-turbo-det}, the upper data flow path consists of three parallel 1D MMSE linear equalizers, the LAIP detector and the first IRA decoder. The lower path includes a 2D MMSE linear equalizer, the BCJR detector and the second IRA decoder. The decoded LLRs from the first IRA decoder in the upper path serve as {\em a priori} information for the BCJR detector in the lower path. The reason for employing different MMSE linear equalizers in the two paths is explained in section~\ref{sec: tdmr_system} below.

Two detection schemes are investigated: center-track detection (based on a single reader) and three-track detection (based on three readers). In three-track detection, the dotted lines in Fig.~\ref{fig: Y-eff-turbo-det} are activated, i.e., all data flows in the system include three tracks, and the BCJR detector is a 2D BCJR. For center-track detection, the LAIP detector passes only its estimate of the central track to the first IRA decoder, which in turn passes its own estimate of this track as {\em a priori} information to a 1D BCJR detector. Using the output of the 3-input/1-output 2D MMSE filter as its read channel input, the 1D BCJR detector forms LLR estimates of the coded bits on the central track and delivers them to the second IRA decoder for the final decision.

\section{LAIP detector for GFP model}
\label{sec: laip}
\begin{figure}[t!]
\centerline{\includegraphics[width=1.5in]{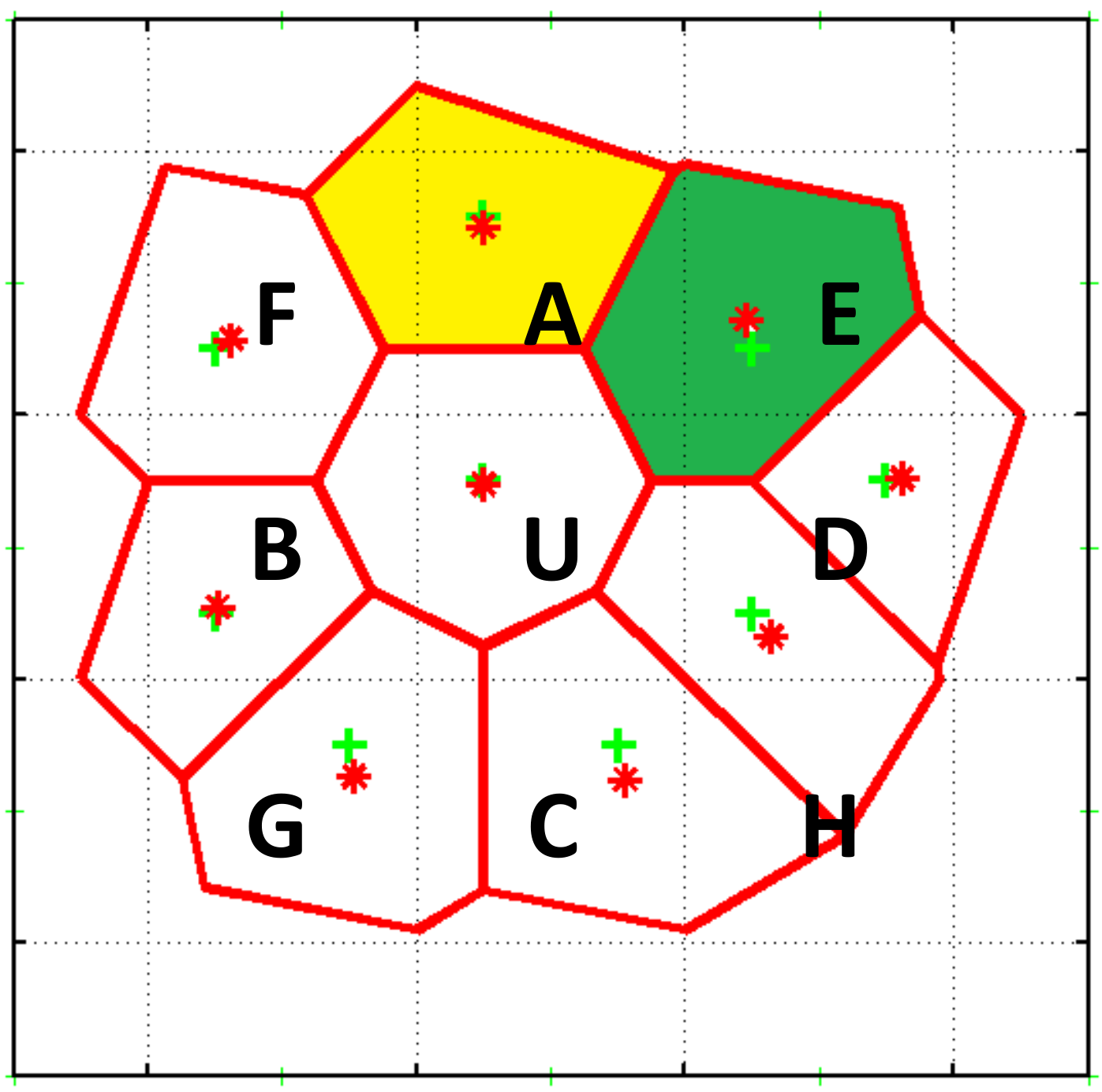}}
\centering
\vspace{-0.1in}
\caption{Areas magnetized by adjacent bits A and E that affect the central coded bit U. The yellow/green portion are the affected area from bit A/E to bit U respectively. }
\label{fig: affected_areas}
\vspace{-0.2in}
\end{figure}

We now describe the LAIP detector block in Fig.~\ref{fig: Y-eff-turbo-det}. 
We assume that the read back value $y$ 
corresponding to a bit U is 
the integral of the magnetizations of all grains contained within the $3 \times 3$ bit cell block centered around bit U, multiplied by the 2D read head impulse response centered at the center of bit cell U. 
The grain magnetization of north/south pole is represented by the values $\{-1, +1\}$, corresponding to a bit value of $\{0, 1\}$, respectively. 
Since each read back value is influenced by grains magnetized by its surrounding eight bit cells, we define 
the integral of the magnetization weighted by the impulse response over the area magnetized by a given surrounding bit as the {\em local area influence} (LAI) on target bit U due to that bit. The LAIs are denoted as $\alpha$; e.g., LAI on bit $U$ from bit $\beta$ is denoted as $\alpha_\beta$.
%
Fig.~\ref{fig: affected_areas} shows portions of central coded bit $U$ affected by vertically and diagonally adjacent bits $A$ and $E$. The yellow portion is the affected area from bit $A$ to bit $U$, and the green is the affected area from bit $E$ to bit $U$. 
The read back value is computed as:
\begin{equation}
y_U = \alpha_U + \alpha_{\mathrm{total}} = \alpha_U + \sum_{i \in \{A,B,...,H\}} \alpha_i,
\label{eq: alpha_total}
\end{equation}  
where $\alpha_U$ denotes the part of reading $y_U$ due only to bit $U$.  Since $\sign(\alpha_U) = \sign(y_U - \alpha_{\mathrm{total}})$, detection comes down to comparing the magnitudes of $y_U$ and $\alpha_{\mathrm{total}}$.
The key idea is that if $y_U$ is significantly greater (less) than $\alpha_{\mathrm{total}}$, then bit U is most likely $+1$ ($-1$). 
However, if $y_U$ and $\alpha_{\mathrm{total}}$ are approximately equal, then it is likely that bit U was overwritten. 
\subsection{LAIP training using GFP data}
\begin{figure}[ht]
\centerline{\includegraphics[width=3.4in]{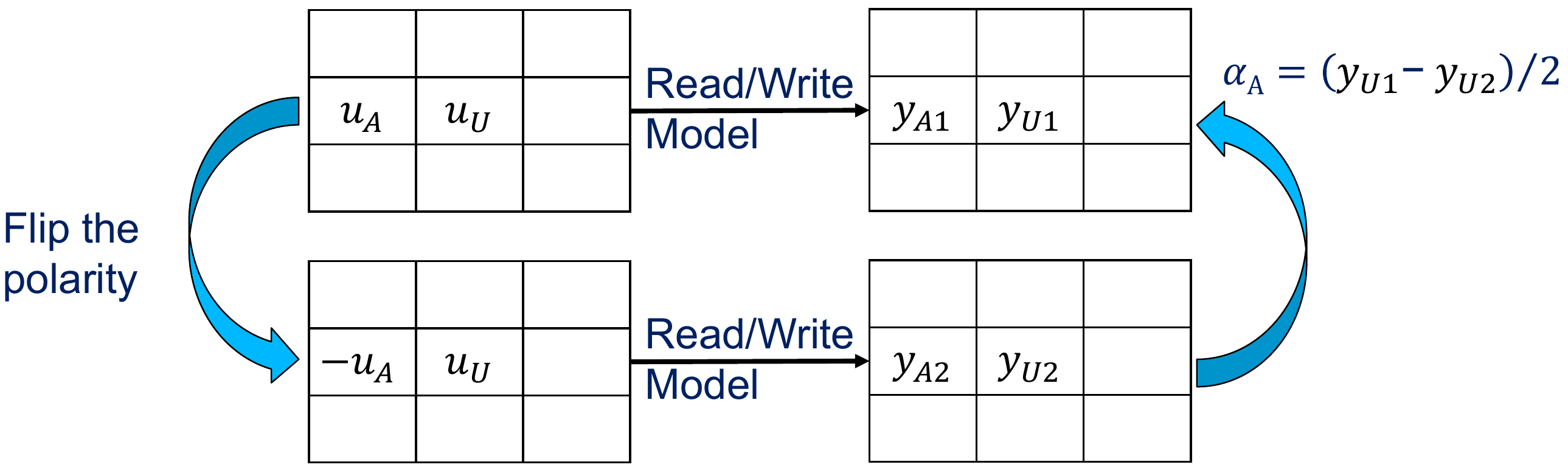}}
\centering
\vspace{-0.1in}
\caption{Designed training pattern approach for GFP model. }
\vspace{-0.1in}
\label{fig: TrainingDesignForGFPModel}
\end{figure}

The dataset from GFP model consists of the input bits $u$ and readback values $y$. In particular, there is no information about the underlying grain boundaries. This is the situation for data collected from a real magnetic disk drive. Thus the LAIs $\alpha$ have to be estimated indirectly, without knowledge of the grain boundaries. Fig.~\ref{fig: TrainingDesignForGFPModel} illustrates our LAI estimation method for a simple case of estimating $\alpha_A$, the influence of bit A on bit U. First, a pair of bits $(u_A, u_U)$ are written, and the sample $y_{U1}$ for target bit U is read back. Then  $(-u_A, u_U)$ is written at the exact same location on the simulated media, and the sample $y_{U2}$ for target bit U is read back again. The difference between the two read back values is caused by the change to bit $A$; therefore, the LAI $\alpha_A$ of $A$ on $U$ can be estimated as $(y_{U1} - y_{U2}) / 2$. In practice, multiple $\alpha_A$ estimates are computed by performing this flipping and subtraction at multiple occurrences of the same pattern in both the downtrack direction and across different readings. The final $\alpha_A$ estimate is then obtained by averaging over these values. This helps achieve a more reliable LAI estimate since in the GFP model grains are flipped through the use of a random number generator and the readings are thus noisy.
%

In order to systematically estimate the LAI corresponding to all possible neighboring bit patterns, we use special training patterns consisting of 512 different $3 \times 3$ patterns of input bits for $(u_A, \ldots, u_H)$. 
For each $3 \times 3$ bit pattern, the $3 \times 4$ bit group comprising the $3 \times 3$ bit pattern followed by a $3 \times 1$ bit random guard band 
column is repeatedly written on and read from the simulated media 10301 times. The guard band ensures that the estimated $\alpha$s are not influenced by the adjacent $3\times3$ bit pattern. These $3 \times 4$ bit groups plus two initial random guard bands and one final guard band make up a total of $3 \times 41207$ bits on three media tracks per pattern file. In the GFP simulation, each of the 512 $3 \times 41207$ 
input files are written on exactly the same underlying simulated grain pattern at exactly the same starting 
location. Each of the 512 input files are read ten times
in order to account for the random grain flipping that occurs during the GFP simulated write process, resulting in a total of 5120 files with $3 \times 41207$ samples in each file.

\subsection{Offline training procedure for LAIP}
To estimate $\alpha_{\mathrm{total}}$ (so as to estimate bit $U$), we train a pre-computed table of $\alpha_{\mathrm{total}}$'s discretized probability mass function (PMF) conditioned on the related read values as shown in Fig.~\ref{fig: affected_areas} ($y_\beta, \beta \in \{A,B,...,H\}$). Each read value $y$ is discretized to 40 bins (with $y$ ranging from around $-2$ to $+2$, the precision should be at least $4 / 40 = 0.1$ to work well experimentally). In practice, to limit the training time and memory storage, we train several different, and much smaller, conditional PMF tables. In real-time estimation, we look them up based on the related read values and convolve them to estimate $\alpha_{\mathrm{total}}$ (explained later in \eqref{eq: P_alphatotal}). The convolution implies an underlying assumption that the smaller PMFs are independent of each other. This independence assumption is only an approximation since the influences from different surrounding bits are correlated due to ISI and ITI. However, the correlations are relatively weaker in the cross track direction (compared to the downtrack direction) due to: (a) the bit length being much smaller than track pitch and (b) correlation introduced by the 1D MMSE equalizer (preceding the LAIP) in the downtrack direction. Therefore, we choose our smaller PMFs to be the combined LAI corresponding to horizontal triplets (F, A, E), (G, C, H) and horizontal pair (B, D). This is an important difference in our LAIP detector design compared to that in \cite{Sun2018}, where the PMFs are trained in terms of pairs of adjacent bits. 

For example, Algorithm~\ref{alg:training} shows the detailed steps for training the  conditional PMF $P(\alpha_{F+A+E} | y_F, y_A, y_E, y_U, u_F, u_A, u_E, u_U)$, corresponding to the influence of the horizontal triplet (F, A, E) on bit U. 
%
\begin{algorithm}
\caption{LAIP offline training procedure}\label{alg:training}
\begin{algorithmic}[1]
\FOR{each 4-tuple $(u_F, u_A, u_E, u_U)$}
\STATE Look up the corresponding readings $(y_F, y_A, y_E, y_U)$ in the training data. 
\STATE Discretize each of the four readings into $40$ bins using Lloyd-Max quantization \cite{Lloyd,Max}.
\STATE From the 512 training patterns, collect all occurrences of the $3 \times 3$ bit pattern $\mathbf{A}_+$ that contains $(u_F, u_A, u_E)$ and their corresponding flipped version $\mathbf{A}_-$ that flips bits $(F, A, E)$, (i.e., contains $(-u_F, -u_A, -u_E)$).
\STATE Look up the readings $y_{U+}, y_{U-}$ in the training data corresponding to $\mathbf{A}_+, \mathbf{A}_-$.
\STATE Compute the LAI $\alpha_{F+A+E}$ for all occurrences as described in Fig.~\ref{fig: TrainingDesignForGFPModel}.
\STATE Compute the average value of $\alpha_{F+A+E}$ over all occurrences.
\STATE Discretize the average $\alpha_{F+A+E}$ into $41$ equal bins spanning values $-2$ to $2$. 
\STATE Save the relevant indices and the frequency count for $P(\alpha_{F+A+E} | y_F, y_A, y_E, y_U, u_F, u_A, u_E, u_U)$, after normalizing the counts so that they sum to 1.
\ENDFOR
\end{algorithmic}
\end{algorithm}
%
%
This PMF is used when {\em a priori} information from a previous LAIP iteration is available; in the initial LAIP step, such {\em a priori} information is unavailable and we train and use the PMF $P(\alpha_{F + A + E} | y_F, y_A, y_E, y_U)$ instead. In the cases when the count (in Algorithm~\ref{alg:training}, step 9) is zero, we simply assign probability 1 to the middle 21st bin of the affected area which corresponds to value 0, and probability 0 to all the other bins. The number of bins for $\alpha$ is chosen to be an odd number (41 in our experiments) so that there is a unique middle bin. Our choice of $40$ bins for $y$ and $41$ bins for $\alpha$ were made in order to balance simulation time and detector performance. PMFs $P(\alpha_{B+D} | y_B, y_D, y_U, u_B, u_D, u_U)$ and $P(\alpha_{G+C+H} | y_G, y_C, y_H, y_U, u_G, u_C, u_H, u_U)$ are trained in a similar manner. 

When the central bit U is on the top or bottom row (or leftmost or rightmost column) some of the bits in its $3 \times 3$ neighborhood will be on the boundary, where the input bits are known (based on a known boundary condition), but there will be no read values $y$ available for the boundary bits. We train additional PMF tables to handle such cases. These tables include:
\begin{itemize}
    \item $P(\alpha_{F+A+E} | y_U, u_F, u_A, u_E, u_U)$ (when bit $U$ is in the top row, bits $F,A,E$ are known boundary bits);
    \item $P(\alpha_{G+C+H} | y_U, u_G, u_C, u_H, u_U)$ (when bit $U$ is in the bottom row, bits $G,C,H$ are known boundary bits);
    \item   $P(\alpha_{F+A+E} | y_A, y_E, y_U, u_F, u_A, u_E, u_U)$, 
    \item $P(\alpha_{B+D} | y_D, y_U, u_B, u_D, u_U)$ 
    \item and $P(\alpha_{G+C+H} | y_C, y_H, y_U, u_G, u_C, u_H, u_U)$ (when bit $U$ is in the first column, bit $F, B, G$ are known boundary bits);
    \item $P(\alpha_{F+A+E} | y_F, y_A, y_U, u_F, u_A, u_E, u_U)$,
    \item $P(\alpha_{B+D} | y_B, y_U, u_B, u_D, u_U)$,and
    \item $P(\alpha_{G+C+H} | y_G, y_C, y_U, u_G, u_C, u_H, u_U)$ (when bit $U$ is in the last column, bit $E, D, H$ are known boundary bits).
\end{itemize}

The PMFs should, in theory, be anti-symmetric with respect to the conditioning
$y$ variables; e.g., $P(\alpha_{F+A+E} | y_F, y_A, y_E, y_U) = P(-\alpha_{F+A+E} | -y_F, -y_A, -y_E, -y_U)$;
in practice, this is not always true for the trained PMFs. To enforce anti-symmetry and thus improve the performance of the LAIP detector, we first use 
$3 \times 3 \times 3$ symmetric spatial filters to smooth the PMFs and then force the anti-symmetry property by 
assigning 
$(P(\alpha_{F+A+E} | y_F, y_A, y_E, y_U) + P(-\alpha_{F+A+E} | -y_F, -y_A, -y_E, -y_U))/2$
to both $P(\alpha_{F+A+E} | y_F, y_A, y_E, y_U)$ and $P(-\alpha_{F+A+E} | -y_F, -y_A, -y_E, -y_U)$.

\subsection{Real-time LAIP detection}
The discrete PMF $P(\alpha_{\mathrm{total}})$ is computed in real-time as follows: the LAIP detector looks up the appropriate PMFs according to the surrounding read back values (and input bits, if {\em a priori} information is available) and then computes
\begin{equation}
\begin{split}
P(\alpha_{\mathrm{total}}) &= P(\alpha_{F+A+E} | y_F, y_A, y_E, y_U) \\
&* P(\alpha_{B+D} | y_B, y_D, y_U) \\ 
&* P(\alpha_{G+C+H} | y_G, y_C, y_H, y_U),
\end{split}
\label{eq: P_alphatotal}
\end{equation}  
where $*$ indicates discrete 1D convolution of the relevant conditional PMFs. 
In the first LAIP iteration, the PMFs in \eqref{eq: P_alphatotal} are exactly read from the pre-stored tables.
In later iterations, {\em a priori} information is available from previous LAIP loops, and the LAIP detector computes conditional PMFs by marginalizing over the input bits. For example, consider the triplet $(F,A,E)$: 
\begin{equation}
\begin{split}
     &P(\alpha_{F+A+E} | y_F, y_A, y_E, y_U) = \\
     &\sum_{u_F,u_A,u_E,u_U} [P(u_F,u_A,u_E,u_U) \\
     &\times P(\alpha_{F+A+E} | y_F, y_A, y_E, y_U, u_F, u_A, u_E, u_U)] ,
\label{eq: P_of_alpha_beta_U}
\end{split}
\end{equation}
where $P(u_F,u_A, u_E, u_U) = \prod_{i \in \{F,A,E,U\}} P_{\mathrm{in}}(u_i)$, and $P_{\mathrm{in}}(u_i), i \in \{F,A,E,U\}$ denotes the incoming {\em a priori} probabilities from the previous LAIP iteration.   

With an estimate of $\alpha_{\mathrm{total}}$, the binary output LLR for the coded bit $U$ can be computed as follows:
\begin{equation}
LLR(U) = \log\frac{P(\alpha_{\mathrm{total}} < y_{U}) + P_{\mathrm{ovw}}/2}{P(\alpha_{\mathrm{total}} > y_{U}) +  P_{\mathrm{ovw}}/2},
\label{eq: LLR_Yeff_over}
\end{equation}
where $P_{\mathrm{ovw}}$ is the probability that bit U is overwritten, i.e., $P_{\mathrm{ovw}} = P(y(\alpha_{\mathrm{total}}) = y(y_{U}))$.
If we ignore the overwritten cases $P_{\mathrm{ovw}}$, the LLR in \eqref{eq: LLR_Yeff_over} is
approximately equal to the {\em a posteriori probability} (APP) LLR 
$\log[P(U=+1|y_{U},y_A,\ldots,y_H)/P(U=-1|y_{U},y_A,\ldots,y_H)]$. When overwrite occurs, it is reasonable to assign half of the overwritten probability to $P(U=+1|y_{U},y_A,\ldots,y_H)$ and the other half to $P(U=-1|y_{U},y_A,\ldots,y_H)$. 
The real-time LAIP detection procedure is summarized in Algorithm~\ref{alg:detection}.
Since the real-time detection only involves pre-stored-table lookup and linear convolution, it is computationally efficient.

\begin{algorithm}
\caption{LAIP real-time detection procedure}\label{alg:detection}
\begin{algorithmic}[1]
\FOR{each target bit U in the three-track test data}
\STATE Collect the nine readings $y_U$, $y_A, \ldots, y_H$ in the $3 \times 3$ bit pattern as illustrated in Fig.~\ref{fig: affected_areas}.
\STATE Look up the bin index for each reading from the pre-stored Lloyd-Max bin-boundaries.
\IF{initial LAIP loop}
\STATE Look up the following PMFs for the triplets/pair in the pre-stored PMF tables: $P(\alpha_{F+A+E} | y_F, y_A, y_E, y_U)$, $P(\alpha_{B+D} | y_B, y_D, y_U)$, $P(\alpha_{G+C+H} | y_G, y_C, y_H, y_U)$.
\ELSE
\STATE Look up the following PMFs in the pre-stored PMF tables : 
\begin{equation*}
\begin{split}
    &P(\alpha_{F+A+E} | y_F, y_A, y_E, y_U, u_F, u_A, u_E, u_U), \\
    &P(\alpha_{B+D} | y_B, y_D, y_U, u_B, u_D, u_U), \\
    &P(\alpha_{G+C+H} | y_G, y_C, y_H, y_U, u_G, u_C, u_H, u_U).
\end{split}
\end{equation*}
Marginalize them over the respective input bits using \eqref{eq: P_of_alpha_beta_U} to obtain PMFs conditioned only on read values. 
\ENDIF
\STATE Compute $P(\alpha_{\mathrm{total}})$ using \eqref{eq: P_alphatotal}.
\STATE Compute $LLR(U)$ using \eqref{eq: LLR_Yeff_over}.
\ENDFOR
\end{algorithmic}
\end{algorithm}

\section{LAIP/BCJR TDMR Detection}
\label{sec: tdmr_system}
Since the LAIP detector only takes into account grain-bit interactions inside a local $3 \times 3$ region around
each target bit, we pass its estimated LLRs as {\em a priori} information to a BCJR detector with $N$ trellis stages to include the effect of longer range interactions caused by ISI and ITI (for three-track detection). This section describes the combined LAIP-BCJR system for TDMR detection.

As shown in the block diagram in Fig.~\ref{fig: Y-eff-turbo-det}, the system accepts three tracks of GFP waveforms as input and contains two paths of data processing. 
In the first path, the GFP waveforms enter a 1D MMSE equalizer followed by the LAIP detector. 
This MMSE equalizer employs three identical 1D filters (of size $1 \times 15$) for all three tracks of readings. The reason for choosing the same filter for all three tracks is that the LAIP detector requires additivity when estimating and summing the LAIs, i.e., $\alpha_\beta = (y_{U1} - y_{U2})/2$ and $y_U = {\sum_{\beta \in {A,B,...,H}} \alpha_{\beta}} + \alpha_U$. Additivity is maintained when the same filter $\mathbf{h}$ processes all samples, i.e., $\mathbf{h}*\mathbf{\alpha}_{\beta, k} = (\mathbf{h}*\mathbf{y}_{U1,k} - \mathbf{h}*\mathbf{y}_{U2,k})/2$ and $\mathbf{h}*\mathbf{y}_{U,k} = {\sum_{\beta \in {A,B,...,H}} \mathbf{h}*\mathbf{\alpha}_{\beta,k}} + \mathbf{h}*\mathbf{\alpha}_{U,k}$), where $\mathbf{y}_{U,k}$ denotes the set of $1 \times 15$ readings centered around downtrack position $k$. The filter $\mathbf{h}$ is 1D (of size $1 \times 15$) and not 2D (of size $3 \times 15$). This is because the top (bottom) track has no adjacent track above (below) it with readings for a 2D filter to be applied on. 
In addition, this MMSE equalizer shapes its output to a FR target of $1$, since the LAIP detector inherently assumes binary input bits. Future work could consider more general PR masks for the LAIP. This would require redesigning the LAIP detector to handle M-ary input symbols.

The three-track filtered output from the first MMSE equalizer is passed as input to the LAIP detector. The LAIP detector performs a joint detection on all three rows in five self-loops using Algorithm~\ref{alg:detection}. 
The detection errors based on LAIP output LLRs are significantly reduced during the five self-loops; the number five is found experimentally as the smallest number for the LLRs to converge. Before the LAIP output LLRs are passed into the first IRA decoder, they are fed into a row-wise de-interleaver (not shown in Fig.\ \ref{fig: Y-eff-turbo-det}), because we assume the GFP waveforms are first encoded and then interleaved (as is the case in real HDD waveforms). The IRA decoder then uses row-by-row coset decoding based on known input bits to form its estimate of the coded bits. Its output LLRs are first interleaved and then passed to the BCJR detector in the second path as {\em a priori} information (the interleaver and de-interleaver are not shown in Fig.\ \ref{fig: Y-eff-turbo-det}). 

In the second path of the TDMR detection system, the GFP waveforms flow into a 2D MMSE equalizer followed by the BCJR detector. This 2D MMSE equalizer has a different structure than the 1D MMSE equalizer in the first path. It utilizes a 2D filter (of size $3 \times 15$) that attempts to shape the filtered output either to a 1D (3 tap) PR mask (for center-track-only detection) or to a 2D ($3 \times 3$) PR mask (for three-track detection). The MMSE equalizer is 3-input/1-output for the 1D PR mask and 3-input/3-output for the 2D PR mask; boundary bits are used to generate the upper and lower track outputs. In either scenario, the PR mask is designed according to the method described in \cite{Srinivasa_2015}. Specifically, the 1D mask is a monic mask with tap coefficients
\begin{equation}
\label{eqn: h1D}
h_{1D} =  \begin{bmatrix}
    0.2223 & 1 & 0.2224
  \end{bmatrix}
\end{equation}
and the 2D mask is a monic mask with tap coefficients 
\begin{equation}
h_{2D} = \begin{bmatrix}
    0.0028 & 0.1623 & 0.1417 \\
    0.2795 & 1.0000 & 0.2903 \\
    0.2347 & 0.2684 & 0.0780
  \end{bmatrix}.
\end{equation}
The filtered output from the 2D MMSE equalizer and the {\em a priori} LLRs from the first IRA decoder enter the BCJR detector. The BCJR detector employs a trellis based on the PR mask ISI/ITI channel and jointly estimates coded bit LLRs over all three rows through recursive computation \cite{bcjr}. The BCJR uses the LLRs from the first IRA decoder as {\em a priori} probabilities which multiply the gamma probabilities in calculating its own LLR estimates.The BCJR output LLRs are de-interleaved and then fed into the second row-by-row IRA decoder, whose information bit output LLRs are the final decision of the whole system. In our implementation, both IRA decoders use the same code (as is the case in real HDDs).

Multiplicative LLR weights from the detector to the channel decoder are employed in the system: $\mathbf{w}_1$, from the LAIP detector to the first IRA decoder, and $\mathbf{w}_2$, from the BCJR detector to the second IRA decoder. These weights are chosen to be less than one and account for the fact that the LLRs from the detector are not reliable; their values are optimized based on a semi-exhaustive experimental search to achieve the lowest BER at the channel decoder output. For three-track detection, for $\mathbf{w}_1$ we experimentally arrive at different optimal weights for each of the three tracks, i.e., $\mathbf{w}_1 = (0.5, 0.75, 0.5)$ for track $(1,2,3)$ respectively; for $\mathbf{w}_2$ a single optimal weight is found across the three tracks, i.e., $\mathbf{w}_2 = (0.7,0.7,0.7)$. For center-track detection, weights $w_1 = 0.75$, and $w_2 = 0.7$ are used. In addition, the LAIP output LLRs are found to have larger magnitudes compared to the BCJR output LLR magnitudes; therefore, we clip the LAIP output LLRs before passing them to the first IRA decoder to prevent numerical overflow (clipping threshold of 10 was used in our experiments).  

\section{Simulation Results}
\label{sec: simu_results}
This section presents Monte Carlo simulation results for the LAIP/BCJR TDMR detector system on the GFP waveforms. 
These GFP waveforms have the following properties: track pitch (TP) is 18 nm, bit length (BL) is 11 nm, and grains per coded bit (GPB) is 3.491. 
The media and
read/write parameters of the GFP test data set are identical to those of the designed pattern training data set.
The dimensions of the input bits block are $5 \times 41207$, and of the readings are $3 \times 41207$. The LAIP detector employs five self-loops. 

We report our results in terms of user bits per grain (U/G), calculated as $\text{U/G} = \text{achieved-code-rate} / \text{GPB}$, where ``achieved-code-rate'' is the maximum code rate that yields a final BER of $10^{-5}$ or smaller. The code rate of the systematic IRA decoder is increased by puncturing the parity bits. 

\renewcommand{\arraystretch}{1.5}
\begin{table*}
\caption{Performance of the LAIP/BCJR TDMR detector and a 1D-PDNP BCJR detector on the GFP model data} 
\label{tab:  densitycode}
\begin{center}
\begin{tabular}{|c|c|c|c|c|c|c|}
\hline 
\multirow{2}{*}{\textbf{TDMR Detectors}} &
\multicolumn{1}{p{1.5cm}|}{\centering \textbf{Number of tracks detected}} & 
\multicolumn{1}{p{1.5cm}|}{\centering \textbf{User Bits per Grain}} & \multicolumn{1}{p{1.5cm}|}{\centering \textbf{Code \\ Rate}} & \multicolumn{1}{p{2.0cm}|}{\centering \textbf{Raw Channel BER}} & \multicolumn{1}{p{2.0cm}|}{\centering \textbf{Decoded \\ BER}} & \multicolumn{1}{p{1.5cm}|}{\centering \textbf{Decoded FER}} \\
\hline
1D-PDNP & 1 & 0.1891 & 0.6600 &  0.1853 & 0 (1.1750{\em e}-6) & 0 (0.03) \\
\hline
LAIP/BCJR & 1 & 0.2020 & 0.7050 &  0.1853 & 0 (1.1750{\em e}-6) & 0 (0.03) \\
\hline
LAIP/BCJR & 3 & 0.1914 & 0.6683 &  0.1853 & 0 (3.9167{\em e}-7) & 0 (0.01) \\
\hline
\end{tabular}
\end{center}
\vspace{-5mm}
\end{table*}

\subsection{Comparison between LAIP/BCJR and 1D-PDNP}
\label{subsec: Comp-1D-PDNP}
For comparison, we implemented a standard 1D-MMSE 1D-BCJR/PDNP system \cite[chapt. 33]{vasic} that employs the same 3 tap PR mask \eqref{eqn: h1D} used in the LAIP/BCJR single track detector. This system is based on the following autoregressive model:
\begin{equation}
    n_k(\mathbf{u}_k) = \sum_{i=1}^{L} a_i(\mathbf{u}_k) n_{k-i}(\mathbf{u}_k) + \sigma(\mathbf{u}_k) w_k,
\label{eq: 1D-PDNP}
\end{equation}
where $n_k$ denotes the predicted noise sample at downtrack position $k$, $L$ denotes the predictor memory, $a_i$ are the autoregressive coefficients, $w_k \sim \mathcal{N}(0,1)$ is a time-uncorrelated unit-variance Gaussian sequence, and $\sigma(\mathbf{u}_k)$ is the prediction error standard deviation. The coded data bit pattern $\mathbf{u}_k$ spans the downtrack bits $(k-M, \ldots, k, \ldots, k+\delta)$, leading to a total of $2^{\delta+M+1}$ number of possible patterns. Denoting the ISI length as $I$, the total number of trellis states is $2^{\max(I+L,M)+\delta}$. In our implementation, we choose $L = 4, I = 2, \delta = 1, M = 6$, and thus there are $2^7 = 128$ trellis states.

Table~\ref{tab:  densitycode} shows the simulation results for the proposed LAIP/BCJR TDMR detection system on the GFP model. The 1D-BCJR/PDNP system in the table's first row achieves $0.1891$ U/G, which serves as a comparison baseline. In all cases the total decoded error counts are 0, and we use upper bound estimates with 95\% confidence level (shown inside the parentheses) for the BERs and frame error rates (FERs). These upper bounds are calculated as $3/N$, where $N$ is the total number of decoded bits (for BER) or codewords (for FER) \cite{BER_EST1}. The proposed system achieves $0.2020$ U/G when detecting only the central track (second row in table~\ref{tab:  densitycode}), a $6.8\%$ density increase compared to the baseline 1D-BCJR/PDNP detector. The third row in the table shows that the proposed system achieves $0.1914$ U/G for three-track detection, a  $1.2\%$ density increase compared to the 1D-BCJR/PDNP baseline; in addition, the three-track detection system triples the data throughput compared to 1D-BCJR/PDNP. All results in this section~\ref{sec: simu_results} are under one-shot decoding, except that the 1D-BCJR/PDNP does two iterations with the IRA decoder, in order to make a fair comparison with the LAIP/BCJR detector's two IRA decodings. Further density gains should be achievable when turbo detection is exploited.

\subsection{Comparison between LAIP/BCJR and 2D-PDNP}
\label{subsec: Comp-2D-PDNP}
We also implemented a state-of-art 2D PDNP system as described in \cite{Shi_Barry_2018}. This system jointly detects two tracks by employing the following 2D autoregressive model:
\begin{equation}
    \mathbf{n}_k = \sum_{i=0}^{N_p}\mathbf{P}_i(\mathbf{A}_k)\mathbf{n}_{k-i} + \mathbf{\Lambda}(\mathbf{A}_k)\mathbf{w}_k,
\label{eq: 2D-PDNP}
\end{equation}
where $\mathbf{n}_k$ denotes the $2 \times 1$ vector of predicted noise samples from
both tracks at downtrack position $k$, $N_p$ denotes the predictor memory, $\mathbf{A}_k$ denotes the $2 \times (I+J+1)$ pattern matrix, $\mathbf{P}_i$ denotes the $2 \times 2$ autoregressive model coefficients dependent on $\mathbf{A}_k$, $\mathbf{\Lambda}(\mathbf{A}_k)$ denotes the $2 \times 2$ pattern dependent standard deviation matrix with diagonal elements $\sigma_1(\mathbf{A}_k)$ and
$\sigma_2(\mathbf{A}_k)$, and $\mathbf{w}_k \sim \mathcal{N}(0,\mathbf{I})$ is a $2 \times 1$ spatially and temporally white sequence of Gaussian noise vectors. 
The pattern matrix $\mathbf{A}_k$ can be one of all the $4^{(I+J+1)}$ possible bit patterns on the two tracks 
that span the downtrack samples $(k - J, \ldots, k, \ldots, k+I)$. The total number of required trellis states is thus $4^{(N_p + I + J)}$.

Although 2D-PDNP is incorporated into the Viterbi Algorithm (VA) in \cite{Shi_Barry_2018}, we implemented this 2D-PDNP algorithm with BCJR for fair comparison with our proposed LAIP/BCJR system.
In \cite{Shi_Barry_2018}, data is pre-processed by a 2D-MMSE equalizer before the VA/2D-PDNP detection; a method for joint
design of the 2D-MMSE equalizer and 2D PR target is provided. Specifically, a fractionally spaced equalizer of length 
$N_c = 22$ is designed for use on a micromagnetic data set from Ehime University with two samples per bit \cite{Barry}. In our BCJR/2D-PDNP implementation, we design an MMSE equalizer of length $N_c = 11$ using the
method described in \cite{Shi_Barry_2018}, since our LAIP/BCJR system uses only 1 sample
per bit from the GFP data sets.
In the BCJR/2D-PDNP detection, instead of using the designed PR target, we estimate the pattern-dependent target $s(\mathbf{A}_k)$ by averaging the MMSE filter output readings $\mathbf{y}_k$ associated with pattern $\mathbf{A}_k$, as in \cite{Shi_Barry_2018}.

We compare the output BER of our proposed LAIP/BCJR detector with the above described BCJR/2D-PDNP detector on the same GFP model. Table~\ref{tab: comparison_Shi} summarizes the simulation results. The BERs reported
in Table~\ref{tab: comparison_Shi} are detector-only BERs without channel decoding; in our detection system, this means without the second IRA decoder in Fig.~\ref{fig: Y-eff-turbo-det} that makes the final decoding decision. 
In our implementation of the 2D-PDNP system, the GFP readings are first pre-filtered
by the 2D-MMSE of down-track length $N_c$ designed according to the method in \cite{Shi_Barry_2018}.
The parameters of the 2D-PDNP algorithm (i.e., prediction coefficients, sigmas, and estimated targets) are
trained over forty 41K blocks of the filtered readings and corresponding known input bits on both tracks.

Different combinations of the parameters $N_c, N_p, I, J$ are tested and the performance of the BCJR/2D-PDNP detector is summarized in rows two through four of Table~\ref{tab: comparison_Shi}. The BCJR/2D-PDNP detector gives different BERs on tracks 1 and 2; all parameter
settings give about 10.6\% BER on track 1, but the $N_p = 2$ PDNP gives the lowest BER of
12.48\% on track 2. For comparison, the proposed LAIP/BCJR detector has a BER of 5.82\% on track 1, which is approximately $0.55 \times$ the BER of the BCJR/2D-PDNP system, and a BER of 1.00\% on track 2, which is a factor of $0.08 \times$ the BER of the BCJR/2D-PDNP system. In addition, the proposed LAIP/BCJR detector has a factor of $1.5 \times$ the throughput gain compared to the BCJR/2D-PDNP detector.

\renewcommand{\arraystretch}{1.5}
\begin{table}[ht]
\caption{Detector BER comparison with 2D-PDNP detector from \cite{Shi_Barry_2018}}
\label{tab: comparison_Shi}
\vspace{-5mm}
\begin{center}
\begin{tabular}{|c|c|c|c|c|c|c|c|c|}
\hline
\multirow{2}{0.8cm}{\textbf{Detector}} & 
\multicolumn{1}{p{0.2cm}|}{\centering \textbf{$N_c$}} & \multicolumn{1}{p{0.2cm}|}{\centering \textbf{$N_p$}} & \multicolumn{1}{p{0.2cm}|}{\centering \textbf{$I$}} & \multicolumn{1}{p{0.2cm}|}{\centering \textbf{$J$}} & \multicolumn{1}{p{0.7cm}|}{\centering \textbf{Trellis States}}& \multicolumn{1}{p{0.95cm}|}{\centering \textbf{BER Track 1}} & \multicolumn{1}{p{0.95cm}|}{\centering \textbf{BER Track 2}} \\ 
\hline
LAIP/BCJR & NA & NA & NA & NA & 64 & 5.82\% & 1.00\% \\
\hline
2D-PDNP & 11  & 1 & 1 & 1 & 64 & 10.60\% & 12.57\%  \\
\hline
2D-PDNP  & 23  & 1 & 1 & 1 & 64 & 10.60\% & 12.57\%  \\
\hline
2D-PDNP & 11  & 2 & 1 & 1 & 256 & 10.63\% & 12.48\%  \\
\hline
\end{tabular}
\end{center}
\vspace{-2mm}
\end{table}

The above LAIP/BCJR simulation results assume perfectly known top and bottom boundary bits; in practice, this may not be the case. A more realistic scenario assumes decision feedback on the upper boundary row and simple thresholding of the readback values for the lower boundary row. This corresponds to a case where we start reading from the edge of the magnetic media with known top boundary row and an unknown bottom boundary row (except for readback values), and successively read more rows. Therefore, for detection of each subsequent set of three rows, decision feedback can be used for the top boundary row. Thus, we will have a relatively smaller BER of $10^{-5}$ for the top boundary row and a relatively larger BER of ~$18.5\%$ for the bottom boundary row. Preliminary simulation results in \cite{Sun2018} by our group show a $1.2\%$, $5.1\%$, and $35.7\%$ increase in BER at the output of the LAIP detector on the upper, middle, and lower tracks, respectively. 
Another scenario is where both top and bottom boundary rows are unknown (except for readback values), giving a BER of ~$18.5\%$ for both boundary rows. 
Preliminary simulation results in \cite{Sun2018} by our group show a $13.4\%$,  $5.1\%$ and $36.0\%$ increase in BER at the output of the LAIP detector on the upper, middle and lower tracks, respectively. 
Even with these percentage increases, the proposed LAIP/BCJR detector BERs are still well below those for the BCJR/2D-PDNP detector of \cite{Shi_Barry_2018}. 

\section{Conclusion}
In this paper, we design a LAIP/BCJR detection system for TDMR that achieves density gains of $6.8\%$ for center-track-only detection and $1.2\%$ for three-track detection compared to a conventional BCJR/1D-PDNP detector. The LAIP detector stores PMFs as a LUT during offline training and uses it for real-time detection. Details of the training and detection process are provided. Simulation results also show that the proposed system achieves BCJR BER reduction by factors of $0.55$ and $0.08$ for tracks 1 and 2, respectively, compared to a recently proposed two-track BCJR/2D-PDNP system. All the results reported in this paper are under one-shot decoding. Further density gains and BER reductions are possible with turbo detection.

\section*{Acknowledgment}
This work was supported by NSF grants CCF-1218885 and CCF-1817083, and by the Advanced
Storage Technology Consortium (ASTC).

\bibliographystyle{IEEEtran}
\bibliography{tdmr_ref,zzjour_chen_rev2_dir2disi-refer_corrected,dnn}
\end{document}